\documentclass[letter,twocolumn]{jpsj2}
\usepackage{bm}
\usepackage{graphicx}
\usepackage{amsmath,amsfonts,amsthm,amssymb}

\newcommand{\bG}{\mbox{\boldmath $G$}}

\newcommand{\bq}{\mbox{\boldmath $q$}}
\newcommand{\br}{\mbox{\boldmath $r$}}
\newcommand{\bR}{\mbox{\boldmath $R$}}
\newcommand{\bk}{\mbox{\boldmath $k$}}

\title{{\em Ab initio} Derivation of Low-Energy 
Model for Iron-Based Superconductors LaFeAsO and LaFePO}

\author{Kazuma \textsc{Nakamura}$^{1}$\thanks{Electronic mail: kazuma@solis.t.u-tokyo.ac.jp}, Ryotaro \textsc{Arita}$^{1,2}$, and Masatoshi \textsc{Imada}$^{1,2}$} 

\inst{$^{1}$ Department of Applied Physics, 
 University of Tokyo, 7-3-1 Hongo, Bunkyo-ku, Tokyo 113-8656, Japan \\ 
      $^{2}$ CREST, JST, 7-3-1 Hongo, Bunkyo-ku, Tokyo 113-8656, Japan}

\recdate{\today}

\abst{
Effective Hamiltonians for LaFeAsO and LaFePO are 
derived from the downfolding scheme based on first-principles 
calculations and provide insights for newly discovered 
superconductors in the family of LnFeAsO$_{1-x}$F$_x$, 
Ln = La, Ce, Pr, Nd, Sm, and Gd. Extended Hubbard Hamiltonians 
for five maximally localized Wannier orbitals 
per Fe are constructed dominantly from five-fold degenerate 
iron-3$d$ bands. They contain parameters for effective Coulomb 
and exchange interactions screened by the polarization of 
other electrons away from the Fermi level. 
The onsite Coulomb interaction estimated as 2.2-3.3 eV 
is compared with the transfer integrals between the 
nearest-neighbor Fe-3$d$ Wannier orbitals, 0.2-0.3 eV, 
indicating moderately strong electron correlation.  
The Hund's rule coupling is found to be 0.3-0.6 eV. The derived model 
offers a firm basis for further studies 
on physics of this family of materials.  
The effective models for As and P compounds turn out 
to have very similar screened interactions with slightly narrower 
bandwidth for the As compound. 
}
\kword{%
first-principles calculation, effective Hamiltonian, downfolding, constrained RPA method, LaFeAsO, LaFePO, Oxypnictide, high-temperature superconductivity
}

\begin{document}
\maketitle
 
Recent discovery of a new superconductor, 
LaFeAs(O,F),~\cite{Hosono} has triggered extensive studies on the family 
of layered iron arsenide compounds with 
ZrCuSiAs-type or ThCr$_2$Si$_2$-type structure, 
whose superconducting critical temperature $T_{\rm c}$ 
is now raising up to $\sim$55 K~\cite{Ren}.
A mother compound LaFeAsO shows antiferromagnetic 
order~\cite{Mook} with bad metallic 
transport properties~\cite{Hosono} supporting a significant 
role of electron correlations. The antiferromagnetic ordered 
moment $\sim$0.36 $\mu_B$ 
is unexpectedly 
small, implying large quantum fluctuations arising from 
electron correlations with competing ground states. 
Superconductivity appears when carriers are doped by the 
substitution of F for O~\cite{Hosono}, or the introduction of O 
vacancies~\cite{OVacancy}.  
Many of experimental data support a significant role of electron correlation in realizing 
the superconductivity.~\cite{NMR,muSR,PES1,PES2}
Although the As compounds have 
the maximum $T_{\rm c}\sim$ 56 K, the P
compounds show one order of magnitude lower $T_c$.\cite{LaFePO}

Conventional density-functional calculations with the local density 
approximation (LDA) or the generalized gradient approximation (GGA)
have clarified entangled ten-band structure near 
the Fermi level mainly originating from five-fold degenerate 
iron-3$d$ orbitals contained in each of two iron atoms in the 
unit cell.~\cite{Singh,Hirschfeld,Terakura,Ma,Kuroki}  
For the mother material, the initial LDA calculation~\cite{Singh} 
predicted the nonmagnetic ground state in close proximity 
to a ferromagnetic metal while the recent results show 
antiferromagnetic order.\cite{Hirschfeld,Terakura,Ma} 
In particular, the stripe-type antiferromagnetic order is 
correctly reproduced.\cite{Terakura,Ma} The calculated ordered 
moment obtained so far ranges between 1.2 and 2.6 $\mu_B$, \cite{Hirschfeld,Terakura,Ma,Mazin2}  
in contrast to the tiny ordered moment discussed above.
Broad peak structures of magnetic Lindhard function calculated 
by using the LDA/GGA Fermi surface suggest competitions of several 
different ordering tendencies.\cite{Mazin,Dong,Kuroki,Yildirim}  

Although overall experimental results suggest noticeable correlation 
effects, realistic roles of electron correlations on theoretical 
grounds are not clear. The relevance of the correlation effect 
is under active debates.\cite{Anisimov,Haule,Craco} 
It is imperative to estimate at least the effective Coulomb 
interaction from first principles. Furthermore, 
since all the 3$d$ bands of Fe are as a first look wholly 
involved near the Fermi level, it is important to elucidate 
interplay of orbital degeneracy and electron correlation, 
which can be studied only by a model 
for degenerate bands. A reliable theoretical model derived for 
this family of compounds thus provides us with a firm basis for 
understanding the superconductivity and 
also a starting point for exploring further possibility of 
higher $T_{\rm c}$ compounds. 

In this letter, we present {\em ab initio} low-energy 
effective Hamiltonians of LaFeAsO and LaFePO.  
Implications for the superconductivity in F-doped materials are also discussed.
A reliable downfolding scheme has recently been established which 
has enabled derivation of low-energy effective Hamiltonians 
from {\it ab initio} density-functional calculation 
of real materials.~\cite{Aryasetiawan,Solovyev,Imai}  
The low-energy Hamiltonian is derived after eliminating 
higher-energy degrees of freedom and estimating the renormalization 
effect from the high-energy part onto the bands near the Fermi level. 
The accuracy of the downfolding procedure has been established  
in various cases.\cite{Imai,Otsuka,Nakamura} 
Here, we employ this downfolding scheme in deriving
the extended Hubbard models of LaFeAsO and LaFePO,  
consisting of band dispersion (kinetic energy) of electrons at maximally 
localized Wannier orbitals (MLWOs),~\cite{Ref_MLWF} as well as screened Coulomb and exchange interactions. 
Since the ten-fold Fe-3$d$ bands are basically isolated from other bands, 
we derive effective models for these ten bands. 

The extended ten-band Hubbard model reads
\begin{eqnarray}
&&\mathcal{H}
= \sum_{\sigma} \sum_{{\bm R} {\bm R'}} \sum_{nm}  
  t_{m {\bf R} n {\bf R}'} 
                   a_{n {\bm R}}^{\sigma \dagger} 
                   a_{m {\bm R'}}^{\sigma}   \nonumber \\
&&+ \frac{1}{2} \sum_{\sigma \rho} \sum_{{\bm R} {\bm R'}} \sum_{nm} 
  \biggl\{ U_{m {\bf R} n {\bf R}'} 
                   a_{n {\bm R}}^{\sigma \dagger} 
                   a_{m {\bm R'}}^{\rho \dagger}
                   a_{m {\bm R}'}^{\rho} 
                   a_{n {\bm R}}^{\sigma} \nonumber \\ 
&&+J_{m {\bf R} n {\bf R}'} 
\bigl(\!a_{n {\bm R}}^{\sigma \dagger} 
      \!a_{m {\bm R'}}^{\rho \dagger}
      \!a_{n {\bm R}}^{\rho} 
      \!a_{m {\bm R}'}^{\sigma} 
  \!+\!a_{n {\bm R}}^{\sigma \dagger} 
     \!a_{n {\bm R}}^{\rho \dagger}
     \!a_{m {\bm R}'}^{\rho} 
     \!a_{m {\bm R}'}^{\sigma}\bigr)\! \biggr\}, 
\label{eq:H}                
\end{eqnarray}
where $a_{n {\bm R}}^{\sigma \dagger}$ ($a_{n {\bm R}}^{\sigma}$) 
is a creation (annihilation) operator of an electron with 
spin $\sigma$ in the $n$th MLWO centered on Fe atoms in the unitcell at $\bR$. 
$t_{m {\bf R} n {\bf R}'}$ contains single-particle levels and transfer integrals, given by 
$t_{m {\bf R} n {\bf R}'}
=\langle \phi_{m {\bf R}} | \mathcal{H}_0 |  \phi_{n {\bf R}'} \rangle$  
with $| \phi_{n {\bf R}} \rangle =a_{n {\bm R}}^{\dagger}|0\rangle$
and $\mathcal{H}_0$ being the one-body part of $\mathcal{H}$.
$U_{m {\bf R} n {\bf R}'}$ and 
$J_{m {\bf R} n {\bf R}'}$ are screened Coulomb and exchange 
interactions, respectively, expressed as  
\begin{eqnarray}
U_{m {\bf R} n {\bf R}'}
= \langle \phi_{m {\bf R}} \phi_{m {\bf R}} | W | 
    \phi_{n {\bf R}'} \phi_{n {\bf R}'} \rangle 
\label{eq:U}
\end{eqnarray} 
and 
\begin{eqnarray}
J_{m {\bf R} n {\bf R}'} 
= \langle \phi_{m {\bf R}} \phi_{n {\bf R}'} | W | \phi_{n {\bf R}'} 
\phi_{m {\bf R}} \rangle
\label{eq:J}
\end{eqnarray}
with $W$ being a screened Coulomb interaction.
There are already attempts to estimate
$t_{m {\bf R} n {\bf R}'}$ using MLWO~\cite{Kuroki,Hirschfeld}.
We here focus on an {\em ab initio} derivation of the many-body part 
of $\mathcal{H}$; we estimate the interaction 
parameters in eqs.~(\ref{eq:U}) and (\ref{eq:J}), based on 
a constrained random-phase approximation 
(cRPA)~\cite{Aryasetiawan,Solovyev}.
The cRPA has several advantages over
other methods such as constrained LDA~\cite{cLDA}.
We can precisely exclude screening processes 
among the Fe-3$d$ MLWOs 
being the bases of the effective model. 
(This screening should be considered when we solve the effective model.) 
In addition, we can calculate matrix elements of $W$ as a function 
of ${\bf R}$ and ${\bf R'}$;\cite{Ref_Miyake} 
 i.e., we can obtain onsite and offsite interactions at one time.
The cRPA becomes a good approximation, when the high-energy eliminated bands are well separated from the Fermi level and the screened Coulomb interactions between the high-energy electrons and the low-energy electrons are weak. 
 As is noted above, quantitative accuracies of our downfolding 
including cRPA has already been confirmed in a 
number of examples.\cite{Aryasetiawan,Solovyev,Imai,Otsuka,Ref_Miyake} 
In the present case of the iron compounds, 
the condition above is equally satisfied. 

$U_{m {\bf R} n {\bf R}'}$ 
 of eq.~(\ref{eq:U}) is practically calculated in the reciprocal space by using
a Fourier transform of $W$,
\begin{eqnarray}
 W(\br,\br')=\sum_{{\bf q} {\bf G} {\bf G}'} 
e^{i ({\bf q}+{\bf G}) {\bf r}} 
W_{{\bf G} {\bf G}'} 
(\bq) e^{-i ({\bf q}+{\bf G}') {\bf r'}} \label{eq:W}.
\end{eqnarray} 
Here, $\bG$ is a reciprocal lattice vector 
and $\bq$ is a wave vector in the first Brillouin zone. 
We define $W_{{\bf G} {\bf G}'} (\bq)$ as
\begin{eqnarray*}
W_{{\bf G} {\bf G}'} (\bq) = \frac{4 \pi}{\Omega} \frac{1}{| \bq+\bG |} 
 \epsilon_{{\bf G} {\bf G}'}^{-1}(\bq) \frac{1}{| \bq+\bG'|},
\end{eqnarray*} 
where $\Omega$ is the crystal volume and 
$\epsilon_{{\bf G}{\bf G}'}^{-1}(\bq)$ is the
inverse dielectric matrix which
is related to the irreducible polarizability $\chi$ by 
$\epsilon_{{\bf G}{\bf G}'}(\bq) 
= \delta_{{\bf G}{\bf G}'} - v(\bq+\bG) \chi_{{\bf G}{\bf G}'}(\bq)$,  
where $v(\bq) = 4 \pi/\Omega |\bq|^2$ is the 
bare Coulomb interaction. 
The polarization matrix $\chi_{{\bf G} {\bf G}'}(\bq)$ 
is calculated as
\begin{eqnarray*}
\chi_{{\bf G} {\bf G}'}(\bq)&=&
{\sum_{{\bf k}} \sum_{\alpha \beta}}^\prime 
\langle \psi_{\alpha {\bf k}+{\bf q}} 
| e^{i ({\bf q}+{\bf G}) {\bf r}} 
| \psi_{\beta {\bf k}} \rangle \\
&& \times
\langle \psi_{\beta {\bf k}} 
| e^{-i ({\bf q}+{\bf G}) {\bf r}} | 
\psi_{\alpha {\bf k}+{\bf q}} \rangle 
\frac{ f_{\alpha {\bf k}+{\bf q}} - f_{\beta {\bf k}}} 
{E_{\alpha {\bf k}+{\bf q}} - E_{\beta {\bf k}}}, \label{eq:chi} 
\end{eqnarray*}
where $\{ \psi_{\alpha {\bf k}} \} $ are the Bloch states
and the prime attached to the band sum 
indicates that we exclude the 3$d$-3$d$ band transitions
in the calculation of $\chi$. By inserting eq.~(\ref{eq:W}) 
into eq.~(\ref{eq:U}), we obtain the form of 
\begin{eqnarray}
    U_{m {\bf 0} n {\bf R}} 
= \frac{4 \pi}{\Omega} \sum_{{\bf q} {\bf G} {\bf G}'} e^{-i {\bf q} {\bf R}} 
\rho_{m {\bf q}}(\bG) 
\epsilon_{{\bf G} {\bf G}'}^{-1}(\bq) 
\rho_{n {\bf q}}^{*}(\bG'), 
\label{eq:matrix_U}
\end{eqnarray}
where 
\begin{eqnarray*}
\rho_{n {\bf q}}(\bG) = \frac{1}{N | \bq+\bG | } 
\sum_{{\bf k}}^{N} \langle \tilde{\psi}_{n {\bf k}+{\bf q}} 
| e^{i ({\bf q}+{\bf G}) {\bf r}} | 
\tilde{\psi}_{n {\bf k}} \rangle 
\end{eqnarray*}
with  
$
|\tilde{\psi}_{n {\bf k}}\rangle = 
\sum_{{\bf R}}^{N} 
|\phi_{n {\bf R}}\rangle e^{-i {\bf kR}}.
$
It should be noted here that the quantities 
$\langle \tilde{\psi}_{m {\bf k}+{\bf q}} 
| e^{i ({\bf q}+{\bf G}) {\bf r}} | 
\tilde{\psi}_{n {\bf k}} \rangle$ 
can be easily evaluated with the fast Fourier transformation technique. 
Matrix elements of the bare (or unscreened) Coulomb 
interaction as 
$U_{m {\bf 0} n {\bf R}}^{{\rm bare}} = 
\langle \phi_{m {\bf 0}} \phi_{m {\bf 0}} 
| v | \phi_{n {\bf R}} \phi_{n {\bf R}} \rangle$ 
are calculated with replacing 
$\epsilon_{{\bf G} {\bf G}'}^{-1}(\bq)$ of 
eq.~(\ref{eq:matrix_U}) by $\delta_{{\bf G} {\bf G}'}$. 
The parallel treatment is applied to the derivation of 
screened exchange interactions  
in eq.~(\ref{eq:J}). The result is 
\begin{eqnarray}
J_{m {\bf 0} n {\bf R}} 
= \frac{4 \pi}{\Omega} \sum_{{\bf q} {\bf G} {\bf G}'}
\rho_{mn {\bf R}{\bf q}}(\bG) 
\epsilon_{{\bf G} {\bf G}'}^{-1}(\bq) 
\rho_{mn {\bf R}{\bf q}}^{*}(\bG') 
\label{eq:matrix_J}
\end{eqnarray}
with
\begin{eqnarray*}
\rho_{mn {\bf R} {\bf q}} (\bG) = \frac{1}{ N | \bq+\bG | } 
\sum_{{\bf k}}^{N} e^{-i {\bf k} {\bf R}}  
\langle \tilde{\psi}_{m {\bf k}+{\bf q}} | e^{i ({\bf q}+{\bf G}) {\bf r}} | 
\tilde{\psi}_{n {\bf k}} \rangle. 
\end{eqnarray*}

We implemented this scheme in  
{\em Tokyo Ab initio Program Package}.\cite{Ref_TAPP} 
With this program, electronic-structure calculations with the
GGA exchange-correlation functional \cite{Ref_PBE96}
were performed using a plane-wave basis set and the 
Troullier-Martins norm-conserving pseudopotentials \cite{Ref_PP1} 
in the Kleinman-Bylander representation \cite{Ref_PP2}. 
Our iron pseudopotential was constructed under the reference 
configuration (3$d$)$^{6.0}$(4$s$)$^{1.8}$(4$p$)$^{0.0}$ by 
employing the cutoff radii for the 3$d$, 4$s$, and 4$p$ 
 states at  2.1 Bohr.
The energy cutoff was set to 64 Ry and a 9$\times$9$\times$5 $k$-point 
sampling was employed. 
The experimental crystal-structure data were taken 
from ref. \citen{Mook} for LaFeAsO and ref. \citen{LaFePO} for LaFePO.  
The dielectric matrices were expanded in plane waves 
with an energy cutoff of 64 Ry and the total number 
of bands included in the sum in eq.~(\ref{eq:chi}) was set to 70. 
The sum of $\bk$ in eq.~(\ref{eq:chi}) was evaluated by 
the tetrahedron method. The additional terms in the 
long-wavelength dielectric matrix due to nonlocal 
terms in the pseudopotentials was explicitly considered 
following ref. \citen{Louie}. The point $\bq=\bG={\bf 0}$ 
in eqs.~(\ref{eq:matrix_U}) and (\ref{eq:matrix_J}) 
requires special handling because of the singularity 
in the Coulomb interaction, which was treated in the manner 
described in ref. \citen{Louie}.

\begin{figure}[h]
\centering
\includegraphics[width=0.5\textwidth]{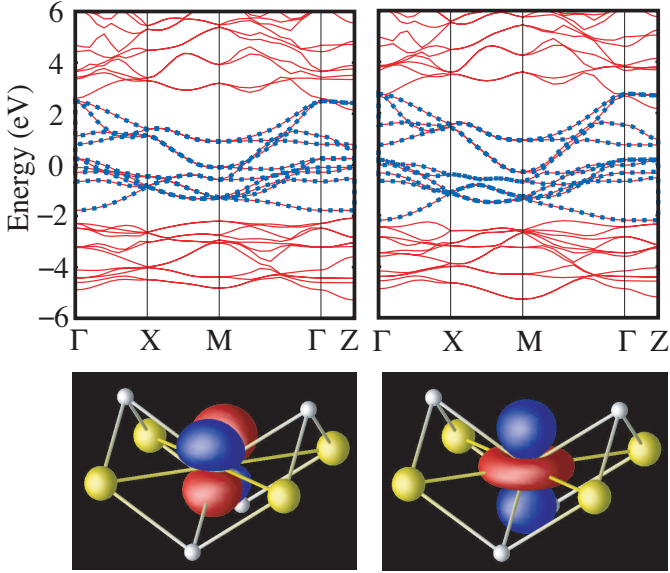}    
\caption{Upper panels: {\em Ab initio} 
band structures of LaFeAsO (left) and LaFePO (right).
Red line and Blue dots are original-GGA and 
Wannier-interpolated bands, respectively. 
The zero of energy is the Fermi level. 
Lower panels: Isosurface contours of $yz$- (left) and 
$z^{\rm 2}$- (right) MLWOs in LaFeAsO. 
The amplitudes of the contour surface are +1.5/$\sqrt{v}$ (blue) 
and $-$1.5/$\sqrt{v}$ (red), where $v$ is the volume of the primitive cell.
Fe and As nuclei are illustrated by yellow and silver spheres, respectively.} 
\label{fig:band_LaOFeAs}
\end{figure}

The upper two panels of Fig.~\ref{fig:band_LaOFeAs} shows 
{\em ab initio} band structures 
of LaFeAsO (left) and LaFePO (right).
We see a good agreement between the original GGA band (red line) 
and the Wannier-interpolated band (blue dots). 
The lower panels visualize our calculated 
MLWOs with the $yz$ (left) and $z^2$ (right) symmetry. 
In principle, the band dispersion thus obtained should be renormalized by 
the interaction between electrons in the Fe-3$d$ and 
eliminated bands in the downfolding procedure. \cite{Aryasetiawan}
In this letter, we assume that this self-energy effect is small 
and thus we employ the same dispersion for the downfolded Hamiltonian.  
The overall band structure of LaFeAsO shows a feature similar 
to that of LaFePO with slightly ($\sim$ 20\%) narrower bandwidth for the As compound.
Around the $\Gamma$ point, in addition to two hole cylinders
with the $d_{yz}$ and $d_{zx}$ characters, 
a 2D-like $d_{x^2-y^2}$ band crosses near the Fermi level for the As compound,
in contrast to a 3D-like $d_{z^2}$band of the P compound. 

\begin{figure}[h]
\centering
\includegraphics[width=0.25\textwidth]{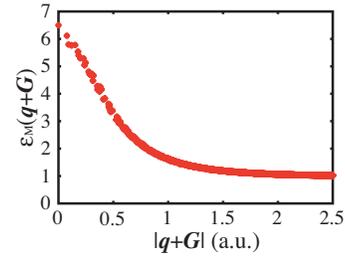}  
\caption{Macroscopic dielectric function of LaFeAsO 
as a function of $|\bq+\bG|$ obtained by the 
constrained RPA method.} 
\label{fig:epsinv_LaOFeAs}
\end{figure}

In Fig.~\ref{fig:epsinv_LaOFeAs}, we plot diagonal 
elements of the macroscopic cRPA dielectric matrix, 
$\epsilon_M (\bq+\bG)=1/\epsilon_{{\bf G}{\bf G}}^{-1}(\bq)$, 
calculated for LaFeAsO, as a function of $|\bq+\bG|$. 
We note that the resulting dielectric 
constant $\epsilon_M^0$ at $\bq+\bG\to0$ 
exhibits a rather high value of 6.3, 
which is compared to that of transition metal oxides such as SrVO$_3$ (6.5). 
It is also interesting to note that 
$\epsilon_M (\bq+\bG)$ does not severely depend
on the direction of $\bq+\bG$, which means that
the screening response is quite isotropic.

\begin{figure}[h]
\centering
\includegraphics[width=0.25\textwidth]{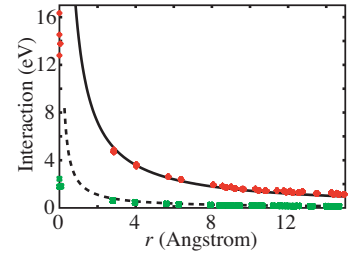}  
\caption{Calculated screened Coulomb interactions of LaFeAsO  
as a function of the distance between the centers of MLWOs. 
Only the interactions between a $d_{xy}$ MLWO at the homecell 
and other MLWOs are plotted.  
The red and green dots represent the bare and screened 
interactions, respectively. The solid and dotted curves denotes 
$1/r$ and $1/(\epsilon_M^0 r)$, respectively.}
\label{fig:W_LaOFeAs}
\end{figure}

Fig.~\ref{fig:W_LaOFeAs} plots matrix elements 
of the screened Coulomb interaction, 
$U_{m {\bf 0} n {\bf R}}$ in eq. (\ref{eq:matrix_U}), 
denoted by green dots, for LaFeAsO as 
a function of the distance between the centers of the MLWOs; 
$r=|\langle \phi_{n {\bf R}}| \br | \phi_{n {\bf R}}\rangle
- \langle \phi_{m {\bf 0}}| \br | \phi_{m {\bf 0}} \rangle |$. 
In this plot, we set $m$ to a $d_{xy}$ MLWO
 and display only the interactions between it and other MLWOs.  
For comparison, we also plot bare Coulomb
interactions $U_{m {\bf 0} n {\bf R}}^{{\rm bare}}$ as red dots, 
 which decay as $1/r$ (solid line)
beyond the nearest-neighbor Fe-Fe distance ($\ge$2.65 $\AA$). 
It is clear that the bare Coulomb interaction
is significantly screened. In addition, 
as expected from Fig.~\ref{fig:epsinv_LaOFeAs}, 
$U_{m {\bf 0} n {\bf R}}$ decays 
as an isotropic function of $1/(\epsilon_M^0 r)$ (dotted line). 
Since the offsite interactions ($U_{m {\bf 0} n {\bf R \neq 0}} $)  
are more than five times smaller than the onsite $U$ values,
we may neglect them in the first step and just start with 
the onsite Hubbard model.
The screened exchange interactions of 
$J_{m {\bf 0} n {\bf R}}$ in eq.~(\ref{eq:matrix_J}) 
are found to decay very quickly; 
the magnitude is nearly zero, except for the onsite values.

We summarize in Table \ref{PARAM} the list of the 
important parameters in the low-energy effective 
Hamiltonian of eq.~(\ref{eq:H}). 
The transfer integral between nearest-neighbor iron 
Wannier orbitals, $t$, is typically 0.2-0.3 eV, whereas 
the onsite screened Coulomb interaction, $U$, 
exhibits 2.2-3.3 eV. 
The strong orbital dependence comes from the fact that each
orbital has different amount of leakage on the neighboring As atoms. 
The onsite exchange interaction (Hund's rule coupling) is 0.3-0.6 eV.
We note that the resulting $U$ and $J$ values 
are smaller than the bare values by the factors 
of about 1/5 and 4/5, respectively.
We note in passing that, for the two-degenerate $d_{yz}$ and $d_{zx}$ orbitals,
 our computed $J_{yz,zx}$ (= 0.45 eV) is close to 
the value estimated from $({U_{yz,yz}-U_{yz,zx}})/2$ (0.50 eV). 
The interaction parameters for the P compound
are found to be very similar; for example, the onsite $U$ values range
from 1.9 eV for $d_{x^2-y^2}$ to 3.3 eV 
for $d_{z^2}$ and $d_{xy}$ as in the case of the As compounds.
\begin{table}[h] 
\caption{List of important parameters (in eV) in the present extended 
Hubbard Hamiltonian in eq.~(\ref{eq:H}). From the top, 
the three 5$\times$5 matrices represent transfer integrals 
between nearest-neighbor sites, onsite screened Coulomb interactions, 
and onsite exchange interactions of LaFeAsO. 
The bottom 5$\times$5 matrix is transfer integrals of LaFePO. 
The onsite energies for the five orbitals of LaFeAsO are 
($\epsilon_{xy}$, $\epsilon_{yz}$, $\epsilon_{z^2}$, $\epsilon_{zx}$, 
 $\epsilon_{x^2-y^2}$) = ($-$0.12, +0.10, $-$0.19, +0.10, +0.16) eV.
} 

\centering 
\begin{tabular}{c@{\,}r@{\,}r@{\,}r@{\,}r@{\,}r@{\ \ \ }r} \hline \hline
  & $t$ (LaFeAsO) & $xy$ & $yz$& $z^2$& $zx$& $x^2\!-\!y^2$ \\
  & & $\phantom{x^2\!-\!y^2}$ & $\phantom{x^2\!-\!y^2}$& $\phantom{x^2\!-\!y^2}$& $\phantom{x^2\!-\!y^2}$& $\phantom{x^2\!-\!y^2}$ \\[-3mm] \hline 
  &   $xy$      &\ $-$0.32& $-$0.25& $-$0.30& $-$0.25&    0.00\\   
  &   $yz$      &\ $-$0.25& $-$0.21& $-$0.08& $-$0.13&    0.18\\ 
  &   $z^2$     &\ $-$0.30& $-$0.08&    0.08& $-$0.08&    0.00\\
  &   $zx$      &\ $-$0.25& $-$0.13& $-$0.08& $-$0.21& $-$0.18\\ 
  &$x^2\!-\!y^2$&\    0.00&    0.18&    0.00& $-$0.18& $-$0.18\\ \hline
\end{tabular}  
\begin{tabular}{c@{\,}r@{\,}r@{\,}r@{\,}r@{\,}r@{\ \ \ }r} \hline
  &$U$(LaFeAsO)        & $xy$ & $yz$& $z^2$& $zx$& $x^2\!-\!y^2$ \\
  & & $\phantom{x^2\!-\!y^2}$ & $\phantom{x^2\!-\!y^2}$& $\phantom{x^2\!-\!y^2}$& $\phantom{x^2\!-\!y^2}$& $\phantom{x^2\!-\!y^2}$ \\[-3mm] \hline 
  &   $xy$        &\ 3.31& 1.95& 1.89& 1.95& 2.09\\   
  &   $yz$        &\ 1.95& 2.77& 2.20& 1.78& 1.67\\ 
  &   $z^2$       &\ 1.89& 2.20& 3.27& 2.20& 1.65\\
  &   $zx$        &\ 1.95& 1.78& 2.20& 2.77& 1.67\\ 
  & $x^2\!-\!y^2$ &\ 2.09& 1.67& 1.65& 1.67& 2.20\\ \hline
\end{tabular} 
\begin{tabular}{c@{\,}r@{\,}r@{\,}r@{\,}r@{\,}r@{\ \ \ }r} \hline 
  &$J$  (LaFeAsO)        & $xy$ & $yz$& $z^2$& $zx$& $x^2\!-\!y^2$ \\
  & & $\phantom{x^2\!-\!y^2}$ & $\phantom{x^2\!-\!y^2}$& $\phantom{x^2\!-\!y^2}$  & $\phantom{x^2\!-\!y^2}$& $\phantom{x^2\!-\!y^2}$ \\[-3mm] \hline 
  &   $xy$      &\  -\ \ & 0.54  & 0.64  & 0.54  & 0.27  \\   
  &   $yz$      &\ 0.54  &  -\ \ & 0.41  & 0.45  & 0.43  \\ 
  &   $z^2$     &\ 0.64  & 0.41  &  -\ \ & 0.41  & 0.50  \\
  &   $zx$      &\ 0.54  & 0.45  & 0.41  &  -\ \ & 0.43  \\ 
  &$x^2\!-\!y^2$&\ 0.27  & 0.43  & 0.50  & 0.43  &  -\ \ \\ \hline 
\end{tabular}
\begin{tabular}{c@{\,}r@{\,}r@{\,}r@{\,}r@{\,}r@{\ \ \ }r} \hline
  & $t$ (LaFePO) & $xy$ & $yz$& $z^2$& $zx$& $x^2\!-\!y^2$ \\
  & & $\phantom{x^2\!-\!y^2}$ & $\phantom{x^2\!-\!y^2}$& $\phantom{x^2\!-\!y^2}$& $\phantom{x^2\!-\!y^2}$& $\phantom{x^2\!-\!y^2}$ \\[-3mm] \hline 
  &   $xy$      &\ $-$0.35& $-$0.31& $-$0.32& $-$0.31&    0.00\\   
  &   $yz$      &\ $-$0.31& $-$0.24& $-$0.04& $-$0.13&    0.18\\ 
  &   $z^2$     &\ $-$0.32& $-$0.13&    0.13& $-$0.04&    0.00\\
  &   $zx$      &\ $-$0.31& $-$0.13& $-$0.04& $-$0.24& $-$0.18\\ 
  &$x^2\!-\!y^2$&\    0.00&    0.18&    0.00& $-$0.18& $-$0.27\\ \hline \hline
\end{tabular} 
\label{PARAM} 
\end{table}
  
The present value of $U$ is substantially smaller than 
the values ($\sim$ 4 eV or larger) employed in model 
studies~\cite{Haule,Anisimov}, which, as a first look, 
suggests that much smaller correlation effects should 
be expected in reality.  However, it should be cautioned 
that the DMFT employed in the literature~\cite{Haule,Anisimov} 
ignoring spatial correlations may largely underestimate 
correlation effects. 
By considering the ratio $U/t\sim 10$ (the value itself is 
comparable to the cuprates) 
and the presence of the five-degenerate orbitals per Fe site, 
electron correlations in these compounds are {\em moderately} strong.
In addition, the present $U$ values have orbital dependence 
with the order of 1 eV, which will generate nonnegligible orbital 
dependence of renormalization factor; it can be a critical parameter 
that affects the Fermi surface in quasipariticle band structure.     
Our downfolded effective Hamiltonian thus poses various constraints on 
modeling of this family of materials and careful analyses 
by accurate low-energy solvers should be required 
for reliable discussions of magnetic and superconducting mechanisms.

The effective models for LaFeAsO and LaFePO are basically 
similar in the band dispersion as well as in the screened 
Coulomb interaction.  A main differences 
is the band width (see Fig.\ref{fig:band_LaOFeAs})
 and this point can also be confirmed 
from $t$ values of LaFePO somewhat larger than 
those of LaFeAsO (see Table~\ref{PARAM}). 
Another difference is a band close to  
the Fermi surface at $\Gamma$ point; i.e., 
whether it has a two-dimensional dispersion with the $d_{x^2-y^2}$ character 
or it has a three-dimensional one with the $d_{z^2}$ character. 
These differences can be systematically understood
in terms of the Fe-Fe and Fe-pnictogen distances~\cite{Vildosola}.  
On the other hand, in our calculations, the interaction parameters do not
exhibit noticeable differences between LaFeAsO and LaFePO.
The mechanism of the superconductivity has to clarify 
how these subtle differences in the one-body and many-body
terms of the effective Hamiltonian
lead to a large difference in the critical temperature, 
26-55 K for As compounds and 4 K for P compounds.

The present model may be used for further  
studies by using low-energy solvers such as 
the dynamical mean-field theory~\cite{DMFT} and 
path-integral renormalization group~\cite{PIRG1,PIRG2}. 
Our ten-band model may also offer a firm starting point 
for further downfolding to derive effective 
models with fewer number of bands near the Fermi level.

This work is supported from MEXT Japan 
under the grant numbers 16076212, 17071003, 17064004, 
19019012, and 19014022. We also thank
the facilities at the Supercomputer Center, Institute for Solid
State Physics, University of Tokyo.
We thank Yoshihide Yoshimoto, Taichi Kosugi, and 
Takashi Miyake for useful discussions.

\end{document}